\documentclass[twocolumn]{aastex62}
\usepackage{hyperref,astrojournals,amsmath,amssymb,mathtools,graphicx,txfonts,microtype,nicefrac,xspace,bm}

\newcommand{\bn}{\begin{enumerate}}
\newcommand{\en}{\end{enumerate}}
\newcommand{\bi}{\begin{itemize}}
\newcommand{\ei}{\end{itemize}}

\def\gtorder{\mathrel{\raise.3ex\hbox{$>$}\mkern-14mu
    \lower0.6ex\hbox{$\sim$}}}
\def\ltorder{\mathrel{\raise.3ex\hbox{$<$}\mkern-14mu
    \lower0.6ex\hbox{$\sim$}}}

\hypersetup{%
  pdftitle={Halo-Bar Coupling},
  pdfauthor={\textcopyright\ authors},
  bookmarksopen=true,
  colorlinks=true,
  linkcolor=black,
  citecolor=black,
  urlcolor=black}

\begin{document}

\title[Halo-Bar Coupling via Resonant Torques]{The Coupling of Galactic Dark Matter Halos with Stellar Bars}

\shorttitle{Halo-Bar Coupling} 

\shortauthors{Collier \& Madigan}

\author{Angela Collier}
\affiliation{JILA and Department of Astrophysical and Planetary Sciences, CU Boulder, Boulder, CO 80309, USA}
\email{angela.collier@colorado.edu}

\author{Ann-Marie Madigan}
\affiliation{JILA and Department of Astrophysical and Planetary Sciences, CU Boulder, Boulder, CO 80309, USA}
\email{annmarie.madigan@colorado.edu}

\begin{abstract}

Resonant torques couple stellar bars to dark matter halos. 
Here we use high-resolution numerical simulations to demonstrate  long-term angular momentum transfer between stellar bars and dark matter orbits of varying orientation. We show that bar-driven reversals of dark matter orbit orientations can play a surprisingly large role in the evolution of the bar pattern speed.

In predominantly prograde (co-rotating) halos, dark matter orbits become trapped in the stellar bar forming a parallel dark matter bar. This dark matter bar reaches more than double the vertical height of the stellar bar.
In halos dominated by retrograde orbits, a dark matter wake forms oriented perpendicular to the stellar bar. These dark matter over-densities provide a novel space to look for dark matter annihilation or decay signals.
We predict that the Milky Way hosts a dark matter bar aligned with the stellar bar as well as a dark matter wake the near-side of which should extend from Galactic center to a galactic longitude of $l \approx 323^\circ$.
\end{abstract}


\section{Introduction}
\label{sec:intro}

\deleted{Stellar bars move disk galaxies in an energetically favorable direction by increasing the density of the inner galaxy and transferring energy and angular momentum outward (for a recent review see \citet{kormendy08}).} 
\added{Stellar bars}
\deleted{They}are drivers of secular evolution, fueling the generation of spiral arms \citep[e.g.,][]{ee85,sellwilk93}, psuedobulges \citep{pfenniger&norman,debattista04,athana05}, and nuclear star formation \citep{combes&elmegreen,kormendy13}.
As a majority of local disk galaxies host stellar bars \citep{erwin18} understanding their dynamics is crucial for appreciating galaxy evolution as a whole.

Stellar bars and the disks in which they form are embedded in massive, responsive dark matter halos which act as reservoirs of angular momentum \citep{sell80,wein85, combes&elmegreen, athana03}.
Recent research has shown that the angular momentum of the halo is an important factor in the bar and disk evolution, affecting bar pattern speed, instability time scales and more \citep{Saha13,pet16,coll18}. Using spectral analysis, \citet{coll19b} discovered an unexpected angular momentum exchange between the stellar bar and retrograde dark matter orbits. They termed this process `orbit reversal’, showing that when retrograde dark matter orbits interact with the stellar bar they can reverse their orientation and become prograde. This allows these orbits to become trapped by the bar, thus increasing the efficiency of angular momentum transfer by the inner Lindblad resonance. 

In this paper we further investigate the role of orbital orientation in the resonant gravitational interactions between the stellar bar and low inclination dark matter halo orbits. We study this problem with high-resolution $N$-body simulations of two-component galaxies containing a stellar disk and dark matter halo (i.e., no gas component). We make this choice to isolate the response of the dark matter halo to the stellar distribution. The results of these numerical experiments can provide insights into how real galaxies evolve.

We present our results as follows. In Section~\ref{halo-bar}, we identify two paths of long-term gravitational coupling between the stellar bar and dark matter halo, one  for prograde (with respect to the stellar disk) dark matter orbits  and the other for retrograde. In Section~\ref{sec:ICs},  we present three high-resolution $N$-body simulations of identical stellar disks inside halos of varying spin. We show how the stellar and dark matter distributions vary in each in Section~\ref{sec:num-results} and link their the differences to angular momentum coupling and bar-driven orbit reversals.
In Section~\ref{obs-cons}, we demonstrate some consequences of these dynamics on observable properties of disk galaxies and on the indirect detection of dark matter. We conclude in Section~\ref{sec:conclusion}.

\begin{figure*}
\centerline{
 \includegraphics[width=\textwidth]{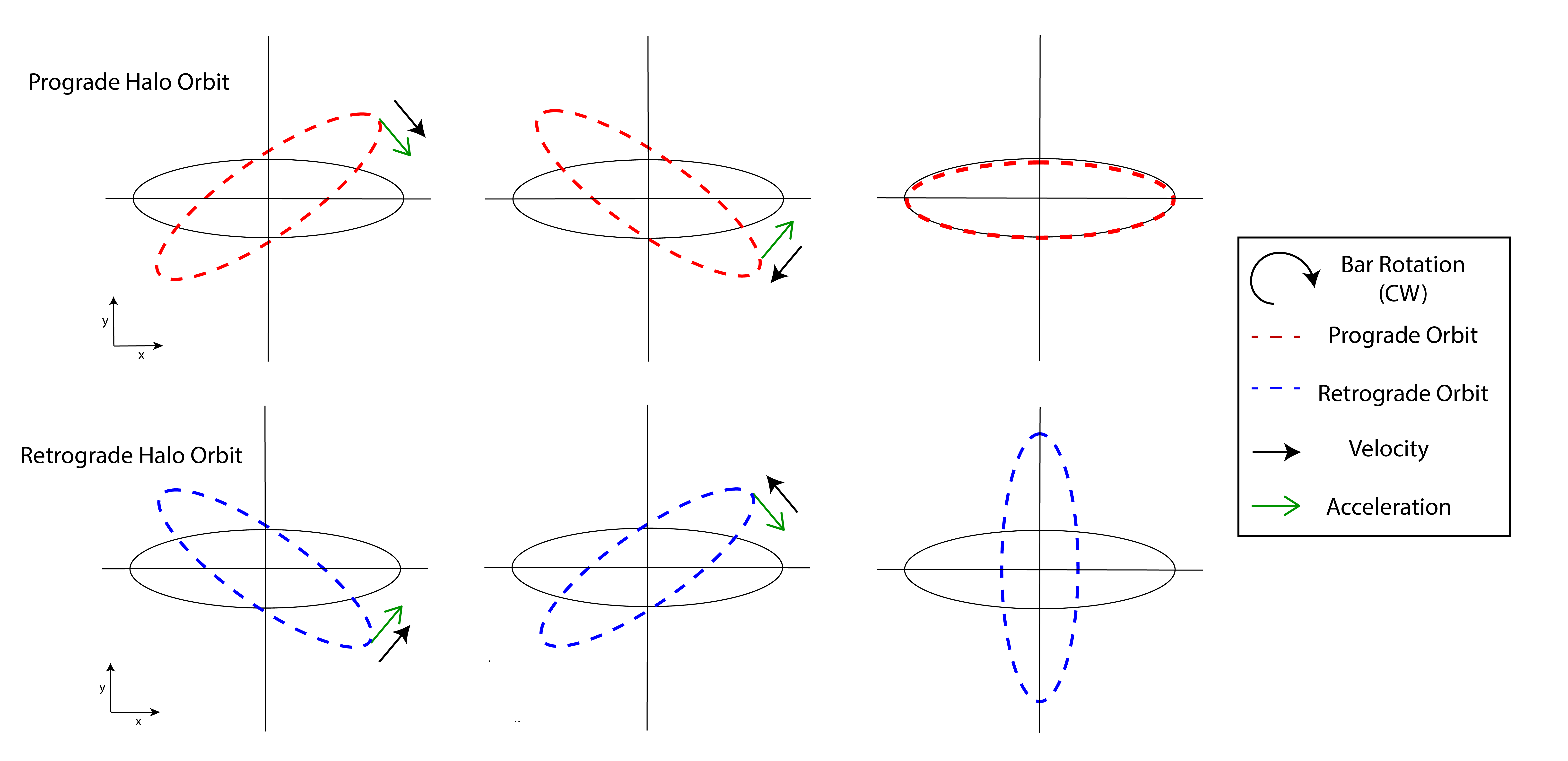}}
\caption{Two paths of dark matter orbit {interaction with} the stellar bar (shown: $x/y$-plane).
 For simplicity, we approximate instantaneous halo orbits as 2D ovals in the frame co-rotating with the bar. Velocity arrows refer to the direction of precession of the dark matter orbits and the acceleration arrows indicate the direction of acceleration of the dark matter orbits due to torque from the bar.
Top: prograde dark matter orbit (dashed red line) being torqued by the bar (solid black line) until aligned. 
Bottom: retrograde dark matter halo orbit (dashed blue line) being torqued by the bar. This orbit will spend more time at right angles to the bar as shown in the rightmost panel. \\
}
\label{fig:orb}
\end{figure*}

\section{Halo-bar coupling}
\label{halo-bar}

{Isolated, disk-less dark matter halos can maintain their kinematic and density distributions over long time scales. However, a strong gravitational perturbation such as a stellar bar will force the halo orbits within the radius of influence of the disk to respond.} We find that the disk's influence on halo orbits is limited to within $\sim 2$ disk scale lengths and $\sim 10$  disk scale heights over a Hubble time.
This means that low inclination halo orbits within the radius of the disk will be most strongly modified by the bar potential.
As non-rotating halos have both retrograde and prograde components within this radius, we discuss how halo orbits of both orientations interact with the stellar bar.

\subsection{Prograde halo orbit trapping}
\label{halo-orbits}

A low inclination dark matter halo particle on a prograde orbit may be trapped by the bar potential in the same manner as a stellar orbit in the disk.
This trapping occurs as follows: we imagine a particle on an elliptical orbit with its major axis slightly ahead of the bar in azimuthal angle and precession frequency equal to the bar's pattern speed.  
The orbit is negatively torqued by the bar and its precession rate slows until it is overtaken by the bar. As it falls behind the bar it is positively torqued and speeds up until it rejoins the bar. Trapped orbits oscillate back and forth across the bar, strengthening its gravitational potential \citep{LB1979}.

Trapping of prograde orbits is shown schematically in the top row of Figure \ref{fig:orb} wherein we have approximated the instantaneous dark matter orbit as a 2D oval.
The trapping of prograde dark matter orbits forms a ``shadow'' dark matter bar, first identified by \citet{pet16}. The dark matter is aligned with the stellar bar and rotates with the same frequency.
{Dark matter orbits that become trapped in the bar resemble $\textrm{x}_1$ orbits (e.g., \citet{conto80}) in the plane of the disk but they retain their three dimensional structure. This results in a dark matter bar that is thicker than the stellar disk.}

\subsection{Retrograde halo orbit wake}
\label{retro-halo-wake}

A retrograde dark matter halo orbit cannot simply align itself with the bar.
If it lies ahead of the bar, it experiences a positive torque (with respect to its angular momentum) which increases the magnitude of its angular momentum. This increases its precession rate (retrograde with respect to the bar) such that it sweeps quickly through the bar, emerging on the opposite side.
Now lagging the bar, it feels a strong negative torque slowing it down in precession. Hence, after a secular timescale, the density of retrograde dark matter orbits will be enhanced at right angles to the bar.
The formation of an over-density or wake of retrograde orbits perpendicular to the bar is shown schematically in the bottom row of Figure \ref{fig:orb} wherein we have approximated an instantaneous dark matter orbit as a 2D oval.
{We emphasize that these retrograde orbits are not trapped in the perpendicular wake; their major axes circulate in azimuthal angle but precess more slowly perpendicular to the bar in response to the bar's resonant torque.}
{The orbits that make up the wake are also distinct from the periodic $\textrm{x}_4$ retrograde family of nearly circular orbits in the stellar disk.}
Even though individual orbits precess with retrograde motion, the newly-formed wake rotates prograde with the stellar bar at the same frequency.
The size of the wake increases as the stellar bar grows and gravitationally interacts with more retrograde dark matter orbits at larger radii.

\subsection{Bar-driven orbit orientation reversals}
\label{bar-driven reversals}

The two paths described above lead to over-densities in dark matter parallel and perpendicular to the stellar bar. They can also lead to a reversal in the orbital orientation of a dark matter particle if the torque from the bar anti-parallel to its orbital angular momentum vector is sufficiently strong. This is most likely to occur at very low orbital inclinations \citep[as in hierarchical three-body systems;][]{Li2014a}.
For a prograde orbit, the 180 degree flip to retrograde orientation will happen when the orbit is ahead of the bar in precession/rotation.
For a retrograde orbit, the 180 degree flip to prograde orientation will happen when the orbit is behind the bar.
The orbit reversals can be periodic with numerous orientation flips as the orbit continues to strongly interact with the bar, unless it receives a boost in inclination in the process.

From this simple analysis we predict that the inner dark matter halo will couple to the galactic bar irrespective of the primary rotation direction of the halo. The resulting dynamics of this coupling will be important to \textit{all} dark matter halos containing barred galaxies, regardless of spin, as all galactic halos will host a population of retrograde and prograde dark matter orbits near the plane of the disk. We test this with numerical simulations in the following sections. 

\section{Numerical Simulations}
\label{sec:ICs}

\begin{table}
\begin{centering}
 \begin{tabular}{|p{1.5cm} |p{1.5cm}| p{1.5cm}
|p{1.5cm} |}
 \hline

 & Retrograde Halo & Non-Rotating Halo & Prograde Halo \\
 \hline
 \hline
Name & Ret & NR & Pro \\
 \hline
 $\lambda$ & 0.101 & 0.002 & 0.101 \\
 \hline
 Prograde  Fraction & 0.0 & 0.5 & 1.0 \\
 \hline
\end{tabular}
\label{table:1}
\caption{Simulation models, with model name, halo spin parameter ($\lambda$), and fraction of particles in the dark matter halo with $z$-components of angular momenta aligned with the stellar disk at $t = 0$.}
\end{centering}
\end{table}

Here we present three models, listed in Table 1, of initially identical stellar disks inside dark matter live halos that vary only in rotation. Our fiducial model has a non-rotating (NR) halo with half the dark matter halo orbits orbiting with the disk and half orbiting against the disk. We also run simulations of fully retrograde (Ret) and prograde (Pro) halos for $10$ Gyr.

We model the stellar disks inside spherical Navarro-Frenk-White {inspired} halos \citep[hereafter NFW]{nava96} using the $N$-body part of the tree-particle-mesh Smoothed Particle Hydrodynamics (SPH/$N$-body) code GIZMO \citep{hop15}.  Our code units for mass, distance, and time are  $10^{10}\,M_\odot$, 1\,kpc, and 1\, Gyr. The halo density follows the NFW profile,

\begin{equation}
\rho_{\rm h}(r) = \frac{\rho_{\rm s}\,e^{-(r/r_{\rm t})^2}}{[(r+r_{\rm c})/r_{\rm s}](1+r/r_{\rm s})^2},
\end{equation}
where $\rho(r)$ is the dark matter density in spherical coordinates, $\rho_{\rm s} \approx 7 \times 10^{10}\,M_\odot$/kpc$^3$ is the fitting density parameter, $r_{\rm s}=9$\,kpc is the characteristic radius, and $r_{\rm c}=1.4$\,kpc is a central density core.  The Gaussian cutoff is applied at $r_{\rm t}=86$\,kpc for the halo. The dark matter halo contains $7.2\times 10^6$ particles and the halo mass is $M_{\rm h} = 6.3\times 10^{11}\,M_\odot$. 

 {The halo velocities are found using an iterative method from \citet{rodio06}, see also \citet{rodio09}. The halo is initialized with the desired density distribution with particle velocities set to zero. The system is then evolved for a short time (0.3 Gyr).
 Each particle in the initial halo is assigned a velocity magnitude from the evolved system using a nearest neighbors scheme. The direction of these velocities are then randomized.}
 This constitutes one iteration. Iterations proceed until the initial velocity distribution is indistinguishable from the evolved velocity distribution. This creates a halo with a non-rotating velocity distribution. Cosmological halos however have a range of spin ($\lambda$).  {The cosmological spin parameter is $\lambda \equiv J_h/\sqrt{2}M_{\rm vir}R_{\rm vir}v_c$ where $J_h$ is the total angular momentum of the dark matter halo, $M_{\rm vir}$ and $R_{\rm vir}$ are the viral mass and radius of the dark matter halo, and $v_c$ is the circular velocity at $R_{\rm vir}$.} Cosmological simulations find that the range of spin parameters can be fit by a lognormal distribution,

\begin{equation}
P(\lambda) = \frac{1}{\lambda\sigma_{\lambda} (2\pi)^{1/2}} \textrm{exp} \bigg[-\frac{\textrm{ln}^2 (\lambda/\lambda_0)}{2\sigma_{\lambda}^2}\bigg],
\end{equation}
where $\lambda_0=0.035\pm 0.005$ and $\sigma_{\lambda}=0.5\pm 0.03$ are the fitting parameters \citep[][]{bull01,hetznecker}.

To create halos of dark matter particles that orbit retrograde/prograde with respect to the disk we reverse the tangential velocities of all prograde/retrograde halo particles in the halo. The new velocity distribution maintains the solution to the Boltzmann equation and does not alter the velocity profile \citep{lynd60,wein85}, so the equilibrium state is preserved. The fully retrograde/prograde halos are within the expected value of halo spin found in cosmological simulations.

\begin{figure}
\centerline{
 \includegraphics[width=0.5\textwidth]{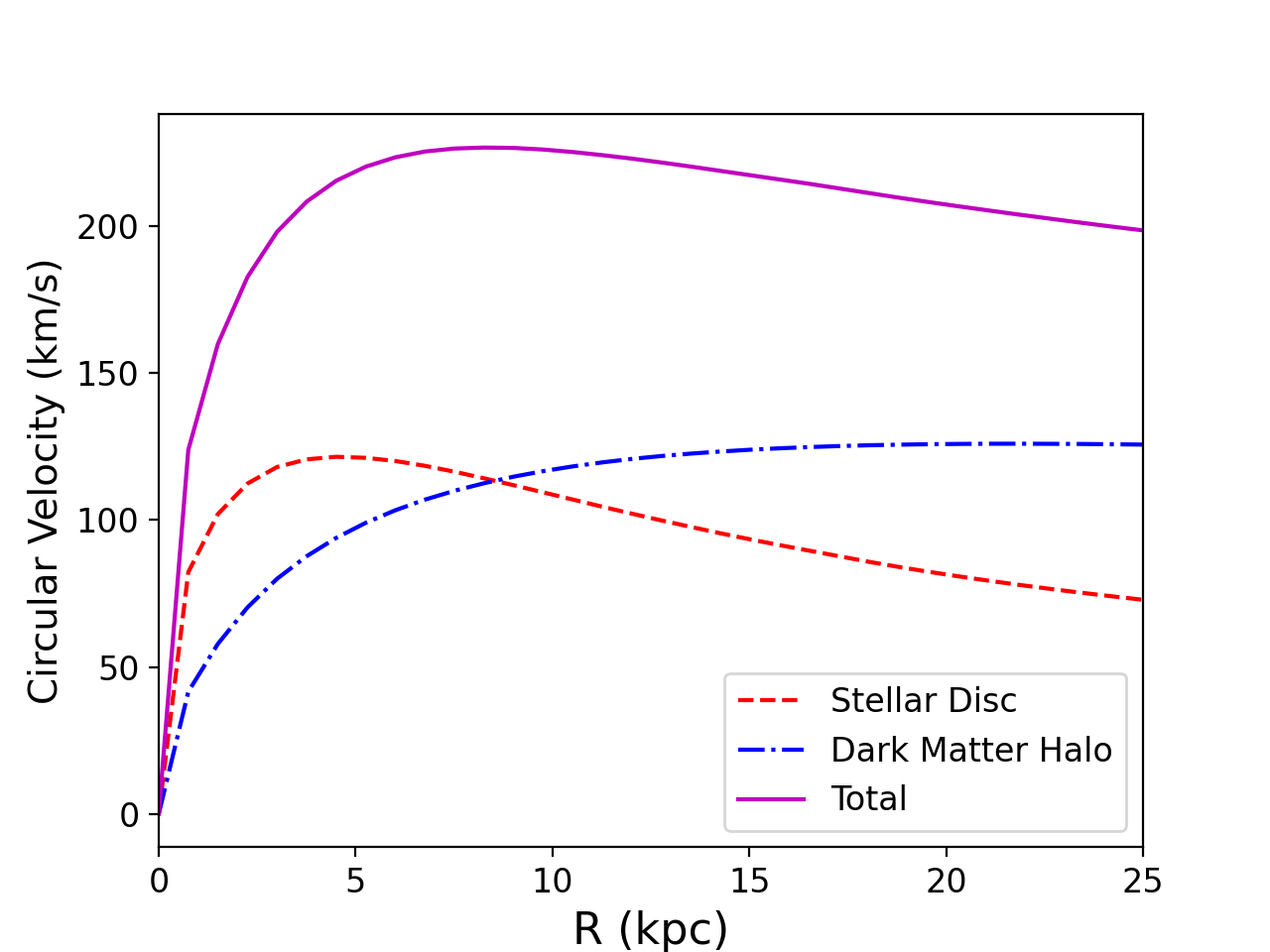}}
\caption{The velocity curve (solid magenta) for our galaxy models, with the stellar disk (red dashed) and dark matter halo (blue dash-dotted) components.}
\label{fig:ic}
\end{figure}

While our halos vary in spin, each simulation begins with an identical disk. The density of the exponential stellar disk is

\begin{eqnarray}
\rho_{\rm d}(R,z) = \left(\frac{M_{\rm d}}{4\pi h^2 z_0}\right)\,{\rm exp}(-R/h)
     \,{\rm sech}^2\left(\frac{z}{z_0}\right),
\end{eqnarray}
where $M_{\rm d}$ is the disk mass, $h=2.85$\,kpc is its radial scale length, and $z_0=0.6$\,kpc is the vertical scale height.  The stellar disk has $N_{\rm d} = 0.8\times 10^6$ particles and the disk mass is $M_{\rm d} = 6.3\times 10^{10}\,M_\odot$. {The radial and vertical dispersion velocities of stellar particles are given by}

\begin{equation}
    \sigma_R(R) = \sigma_{R,0} e^{-R/2h}
\end{equation}

\begin{equation}
    \sigma_z(R) = \sigma_{z,0} e^{-R/2h}
\end{equation}

{where $\sigma_{R,0}$ = 100 \replaced{km/s}{km s$^{-1}$} and $\sigma_{z,0}$ = 80 \replaced{km/s}{km s$^{-1}$}.} 
We plot the velocity curve for our system in Figure \ref{fig:ic}; as the rotation curve is independent of halo spin, this is the same for each model. We note that we have deliberately initialized `hot' disks to delay the buckling instability in our rotating models (as discussed in \citet{athansell86}). 
Furthermore, we expect our gas-free simulations to form bars that are longer than those observed. While this prevents us from making direct comparisons to real, gas-rich galaxies it does not affect the dynamics presented here.

\section{Numerical Results}
\label{sec:num-results}

\begin{figure*}
\centerline{
 \includegraphics[width=\textwidth]{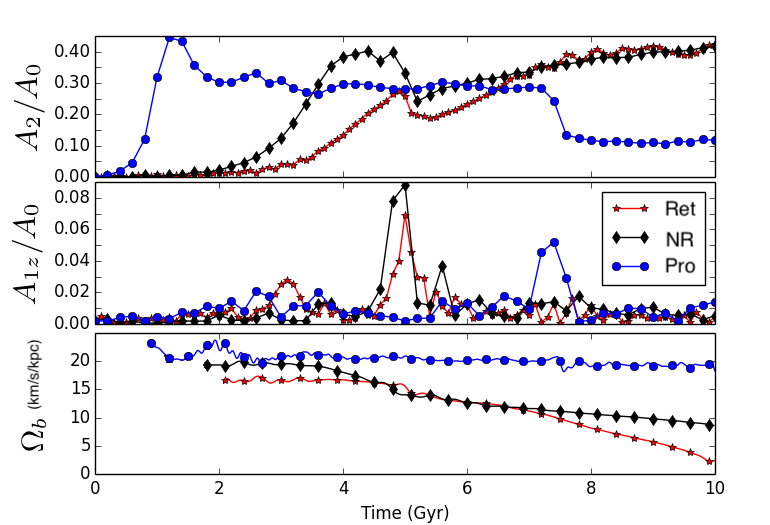}}
\caption{Time evolution of stellar bar parameters for the three galaxy models. Top: Evolution of the bar strength parameter ($A_2/A_0$) in the $x/y$-plane. Middle: Evolution of the ratio of the Fourier $m=1$ mode to the $m=0$ mode ($A_{1z}/A_0$) in the $x/z$-plane. This quantity represents the growth in asymmetry normal to the mid-plane. Bottom: Evolution of bar pattern speed ($\Omega_b$). Models are shown as (blue) circles for the prograde model, (black) diamonds for the non-rotating model, and (red) stars for the retrograde model. }
\label{fig:a2}
\end{figure*}

\subsection{Stellar bar strength, buckling, and pattern speed}

\begin{figure*}[t!]
\centerline{
 \includegraphics[width=\textwidth]{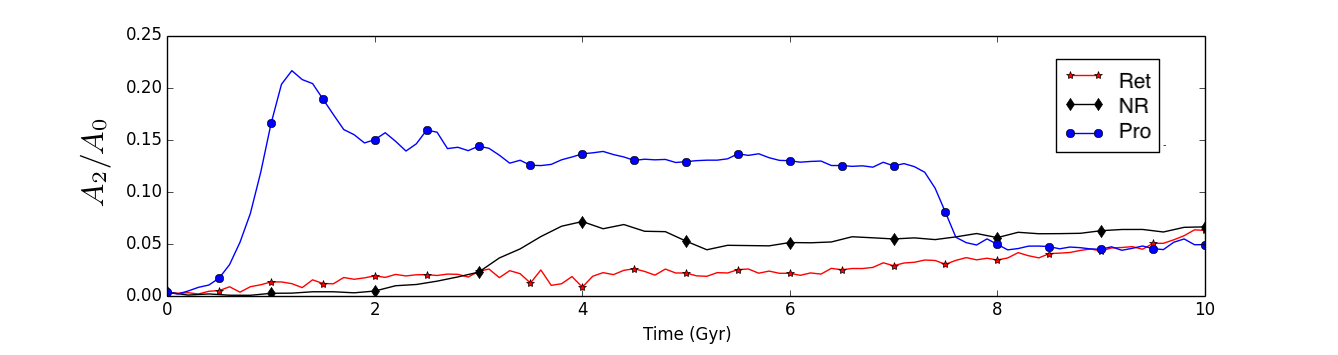}}
\caption{Time evolution of the dark matter bar strength parameter ($A_2/A_0$) in the $x/y$-plane which compares the growth of the Fourier $m=2$ mode to $m=0$ mode the for three galaxy models. This measurement is taken for the slice of the halo $|z|<3$ kpc within $R<12$ kpc. Models are shown as (blue) circles for the prograde model, (black) diamonds for the non-rotating model, and (red) stars for the retrograde model. }
\label{fig:dark matter_a2}
\end{figure*}

We first examine how the evolution of the stellar disks is affected by the varying spins of the dark matter halos. The disks are initially axisymmetric but quickly form bars.  The top panel of Figure \ref{fig:a2} shows the time evolution of the bar strength parameter ($A_2/A_0$) for each model. Here, the strength of the bar is defined by the ratio of the Fourier $m=2$ mode to the $m=0$ mode,

\begin{eqnarray}
\frac{A_2}{A_0} = \frac{1}{A_0}\sum_{j=1}^{N_{\rm d}} m_{\rm j}\,e^{2i\phi_{\rm j}},
\end{eqnarray}
which we get by summing over all stellar particles with $R \leq 12$ kpc and { particle mass $m = m_j$ at azimuthal angle $\phi_j = \arctan{(y/x)}$.}  The prograde (blue circles), non-rotating   (black diamonds), and retrograde (red stars) models all produce strong bars though the time evolution of bar strength for each model is quite different. The strength parameter increases as more orbits get trapped in the bar. Stellar bars are susceptible to a buckling instability due to the increasing radial dispersion velocities associated with the radial bar orbits \citep{hunttoom69}. The bar buckles out of the plane which thickens the disk. The buckling event is associated with a decrease in bar strength \citep{Martinez2004}.

\begin{figure*}[t]
\centerline{
 \includegraphics[width=0.7\textwidth]{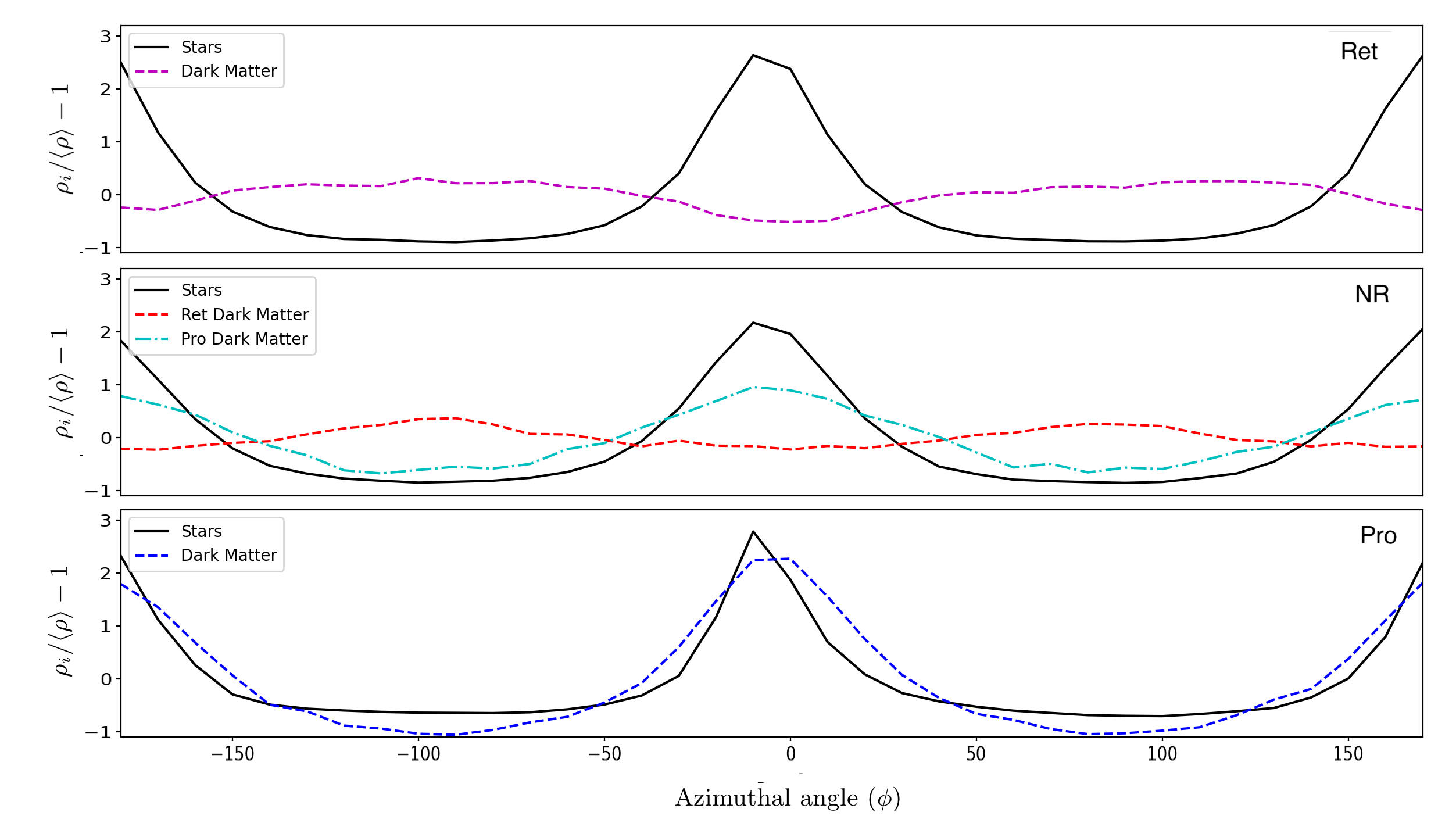}}
\caption{Stellar and dark matter density in each $10^\circ$ sector (i) of an annulus ($4 < R < 6$ kpc \& $\lvert z \rvert < 3$ kpc) compared to the average density in that annulus for each particle species. The stellar disk is shown by the solid black line and the dark matter is shown with dashed lines. Dark matter density fluctuations are lower than stellar densities; each dark matter line is multiplied by three for ease of comparison. These measurements are taken at late times when stellar bars are strong. For the non-rotating and retrograde models, this is at $t=10$ Gyr and for prograde model at $t = 7.8$ Gyr. This plot shows that prograde dark matter orbits align with the stellar bar, while retrograde orbits align perpendicularly. }
\label{fig:dm-angle}
\end{figure*}

The bar forms comparatively early in the prograde simulation, reaching peak strength around $1.8$ Gyr. Shortly after, there is a small reduction in bar strength
followed by $\sim 5$ Gyr of the galaxy hosting a strong bar that is neither growing or decreasing in strength. The bar buckles in this model just before $8$ Gyr and then greatly reduces in strength never to regrow.
The bar instability occurs around $2$ Gyr in the non-rotating model with bar strength increasing exponentially until it reaches its peak. Buckling then reduces the strength of the bar. The bar quickly recovers however, and continues to grow in strength for the rest of the simulation. The retrograde model shows a delayed bar instability. The stellar bar does not reach the same maximum pre-buckling strength as compared to the other two models. However, after buckling, this bar in this model appears strong and grows in strength for the remainder of the simulation.

In the middle panel of Figure \ref{fig:a2}, we plot $A_{1z}/A_0$ which is a measure of asymmetry in the $x/z$-plane. {We calculate this as}

\begin{eqnarray}
\frac{A_{1z}}{A_0} = \frac{1}{A_0}\sum_{j=1}^{N_{\rm d}} m_{\rm j}\,e^{i\phi_{\rm j}},
\end{eqnarray}
which we get by summing over all stellar particles with $R \leq 12$ kpc and particle mass $m = m_j$ at polar angle $\phi_j=\arctan(z/R)$.

This plot is quite noisy with a spike that corresponds to the time of buckling instability in each model. The retrograde and non-rotating models undergo the buckling instability at a similar time while the buckling in the prograde model is delayed. The bar in the prograde halo can trap the maximum number of prograde dark matter halo particles. These dark matter particles have a large spread in inclination which increases the vertical velocity dispersion of the bar. This not only inhibits the buckling instability, it also causes a weaker buckling as compared to the models without a large prograde component. The gentle buckling significantly weakens the bar as described in \citet{coll20}.

The bar slows down as it brakes against the outer disk and halo. The pattern speed acts as a measure of torque (change in angular momentum) experienced by the stellar bar. The pattern speed ($\Omega_b$) is plotted for each simulation in the bottom panel of Figure \ref{fig:a2}.  The retrograde and non-rotating models host bars that are strongly torqued which can be seen in the secular slow-down in the pattern speed. Conversely, the stellar bar in the prograde model barely slows down which tells us that the bar experiences minimal net torques from the halo and outer disk.

\subsection{Dark matter strength and location}
\label{dark matter_bar_mass_reversal}

In Figure \ref{fig:dark matter_a2} we show the bar strength parameter for the dark matter particles in our models. {We limit our measurement to $R<12$ kpc and a thin slice of the halo $\lvert z \rvert <3$ kpc}. The scale of the $y$-axis on this figure is smaller than in Figure \ref{fig:a2}. One immediately notices that the prograde model hosts a much stronger dark matter bar when compared to the other models. The dark matter bar follows the evolution of the stellar bar: bar instability with exponential growth in strength, weakening during buckling, etc.

\begin{figure}
\centerline{
 \includegraphics[width=0.5\textwidth]{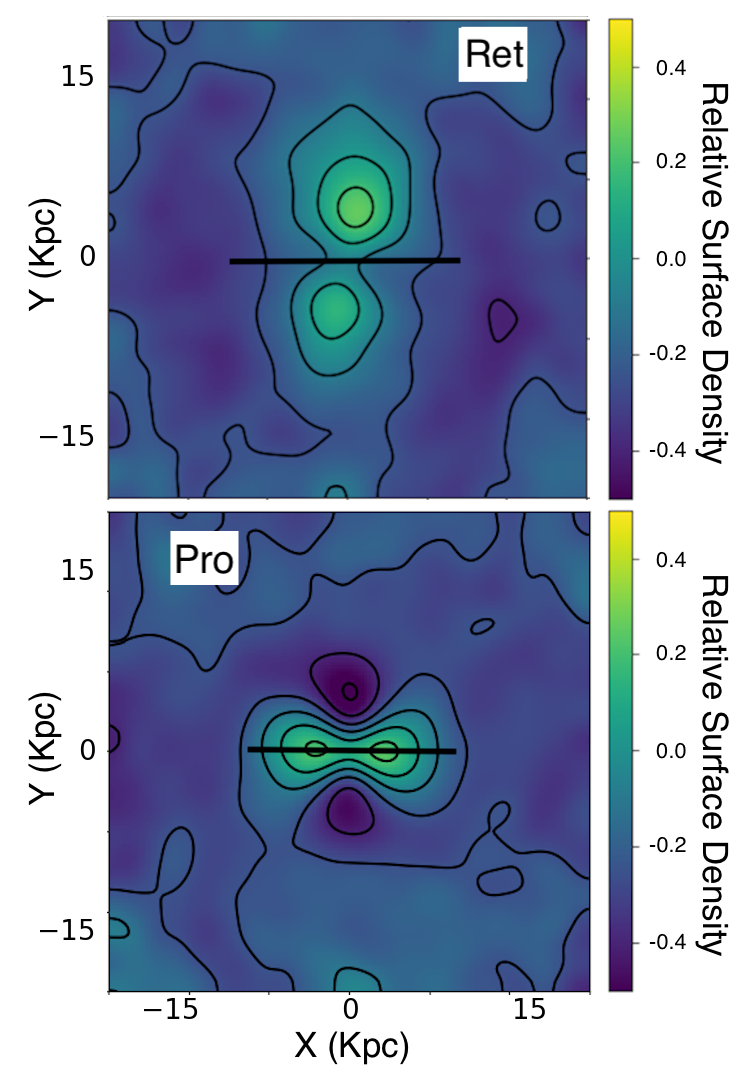}}
\caption{Projected surface density response of dark matter halos in retrograde (top)
versus prograde (bottom) dark matter halos. We show the ratio of the projected dark matter density when the bar is strong compared to the initial density distribution at $t=0$. The contour levels are given in the color palette.
Positions of stellar bars are delineated by the straight horizontal line and they rotate clockwise in this frame.
Note the dark matter bar in the retrograde model is almost perpendicular to the stellar bar.}
\label{fig:wakes}
\end{figure}

{We expect the rotation of the dark matter halo to affect the position of dark matter over-densities with respect to the stellar bar. In Figure \ref{fig:dm-angle} we compare the density of stellar and dark matter in an annulus containing the stellar bar. We include particles in a radial range of $4 \le R \le 6$ kpc and limit the thickness to $\lvert z \rvert < 3$ kpc  for ease of comparison to Figures \ref{fig:dark matter_a2} and \ref{fig:wakes}. We calculate the densities in azimuthal sections of $10^\circ$ and plot them with respect to the average density in the ring. The dark matter lines have been multiplied by a factor of three for ease of comparison.
We position the stellar bar such that its major axis aligns with the $x$-axis. The stellar lines (solid black) resemble saddles with over-densities at $0^\circ$ and $180^\circ$. If the stellar bar had no effect on the dark matter distribution, the dark matter lines on this figure would be flat. This is not the case. In the prograde model (bottom panel) we find that the stellar and dark matter lines strongly overlap. In the retrograde model (top panel) the over-density in dark matter is anti-aligned with the stellar bar. We measure the dark matter distribution for the prograde and retrograde components of the NR model; shown in the middle panel. The prograde component aligns with the stellar bar while the retrograde component is anti-aligned, showing how both the parallel and perpendicular dark matter structures will develop in all halos hosting a stellar bar.}

\begin{figure*}
\centerline{
 \includegraphics[width=\textwidth]{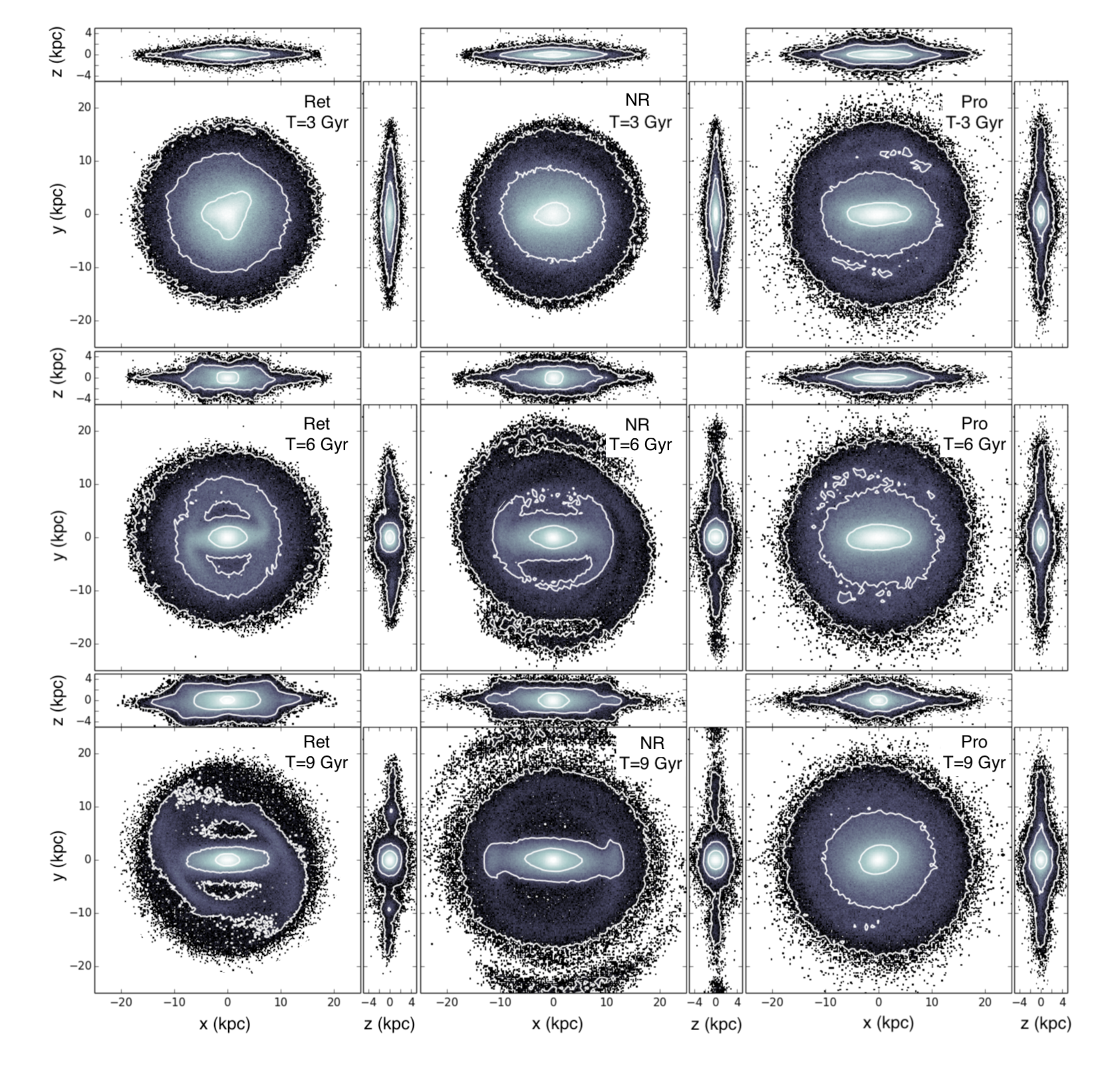}}
\caption{Isodensity contours for the surface density of the stellar disks for the three galaxy models. The left column shows the retrograde model, the middle column shows the non-rotating model, and the right column shows the prograde model. The stellar bars have been aligned along the horizon. Time increases downward. We plot the contours at t= $3,6$ and $9$ Gyr.}
\label{fig:dens}
\end{figure*}

\subsection{Stellar and dark matter density contours}

In addition to affecting the evolution of the stellar bar, the spin of the dark matter halo also affects the kinematics and morphological evolution of the outer disk. We present an isodensity contour time series for face-on and edge-on views of the disks in Figure \ref{fig:dens}. As seen in Figure \ref{fig:a2}, the retrograde and non-rotating models maintain a strong bar until the end of the simulation, while the bar in the prograde model weakens significantly after the delayed buckling instability. This is apparent in the bottom row of Figure \ref{fig:dens}.
In Figure \ref{fig:a2} we show that the retrograde and non-rotating bars slow down for the entire simulation while the prograde model bar stagnates at a constant value.
This stagnation indicates that there is no net torque acting on the outer stellar disk.
This is evident from the absence of spiral arms in the prograde galaxy in the bottom right panel of Figure~\ref{fig:dens}.
Figure \ref{fig:dens} also shows that the stellar disk in the retrograde model does not extend to as large a radius as in the non-rotating model even though the bar is slowing down. If the slow down were due to a negative torque from the outer disk, the outer disk would gain angular momentum and the spiral arms would migrate outwards as in the non-rotating model. Something else is extracting angular momentum from the bar.

In Figure~\ref{fig:wakes} we plot surface density contours of a slice ($z<\lvert 3\rvert$) of the dark matter halos, time-averaged across $\sim0.1$ Gyr, showing how halo orbits align with respect to the stellar bar. The major axis of the stellar bar is indicated by a black line across the horizon of the figure. We find that the prograde model has a strong dark matter bar aligned with the stellar bar, even producing dark matter wakes reminiscent of spiral arms.  In contrast, the over-density of dark matter in the retrograde model appears just offset from perpendicular at $\approx 88^{\circ}$.

\subsection{Angular Momentum Flow}
\label{sec:angmo2}

\begin{figure}
\centerline{
 \includegraphics[width=0.5\textwidth]{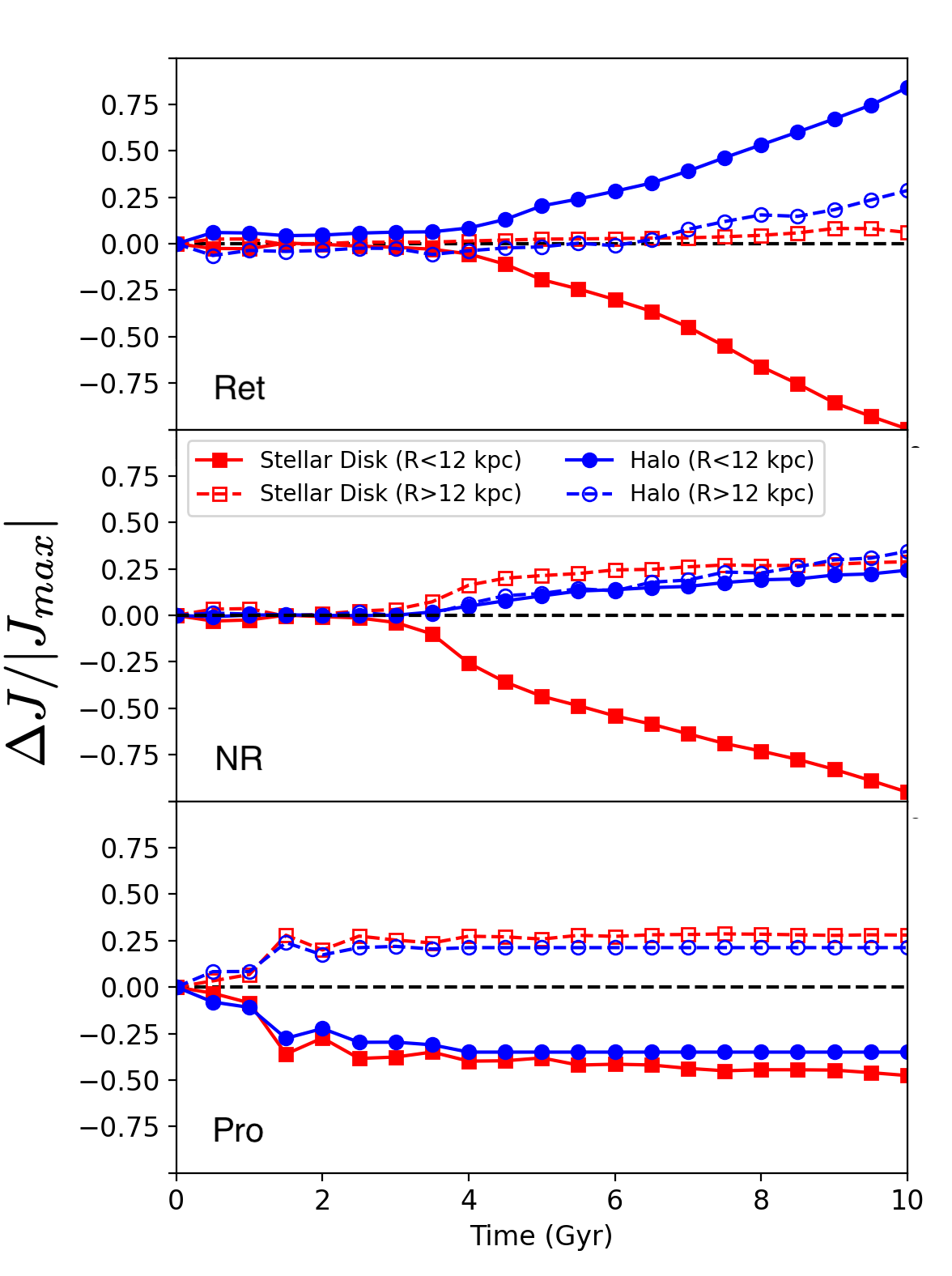}}
\caption{{Time evolution of change in angular momentum, $\Delta J=J(t) - J(t=0)$ normalized to the maximum amount ($J_{max}$) moved by the system. The measurement is made for four discrete regions, inside and outside $R=12$ kpc for each disk and dark matter halo. The disk inside (outside) $R=12$ kpc corresponds to the filled (empty) squares. The dark matter halo inside (outside) $R=12$ kpc corresponds to the filled (empty) circles. The measurement is limited to a cylinder ($\lvert z \rvert < 10$ kpc) as angular momentum does not move significantly beyond this region. The top plot shows results for the retrograde model, the middle plot shows the non-rotating model, and the bottom plot shows the prograde model. }
}
\label{fig:jz2}
\end{figure}

{The stellar bar redistributes angular momentum within the galaxy, the location for which depends on the spin of the dark matter halo. We plot the time evolution of angular momentum for the inner and outer stellar disk and dark matter halo in Figure \ref{fig:jz2}. The measurement is limited to a cylinder ($\lvert z \rvert < 10$ kpc) as angular momentum does not move significantly beyond this region. We distinguish between the inner ($R<12$ kpc) and outer ($R>12$ kpc) disk for ease of comparison with Figures \ref{fig:a2} and \ref{fig:dark matter_a2}. We sum the angular momentum of each particle at each time and subtract that from the value at $t=0$. }

The retrograde model is plotted in the top panel of Figure \ref{fig:jz2}. The inner stellar disk (filled squares) begins to lose angular momentum when the bar appears around $4$ Gyr. In contrast, the halo particles inside the same radius (filled circles) gain angular momentum. The outer disk (open squares) and outer halo (open circles) gain angular momentum as well though to a lesser extent. Thus, the stellar bar in this model mostly moves angular momentum to the inner dark matter halo. As the halo has negative angular momentum at $t = 0$, the gain of positive angular momentum decreases the magnitude of its angular momentum.

{The prograde model is shown in the bottom panel of Figure \ref{fig:jz2}. The inner disk (solid squares) and inner halo (solid circles) move angular momentum outward. The outer disk (open squares) and halo (open circles) gain angular momentum. The dark matter in the inner halo becomes trapped in the stellar bar which prevents it applying torque. Thus, the flow of angular momentum is limited which prevents the expansion of the disk.}

{The middle panel shows the behavior between the two extremes. The non-rotating halo has a prograde and retrograde component allowing angular momentum to move to the outer disk, and inner and outer regions of the dark matter halo.}

\subsection{Disk Radii}

\begin{figure}
\centerline{
 \includegraphics[width=0.5\textwidth]{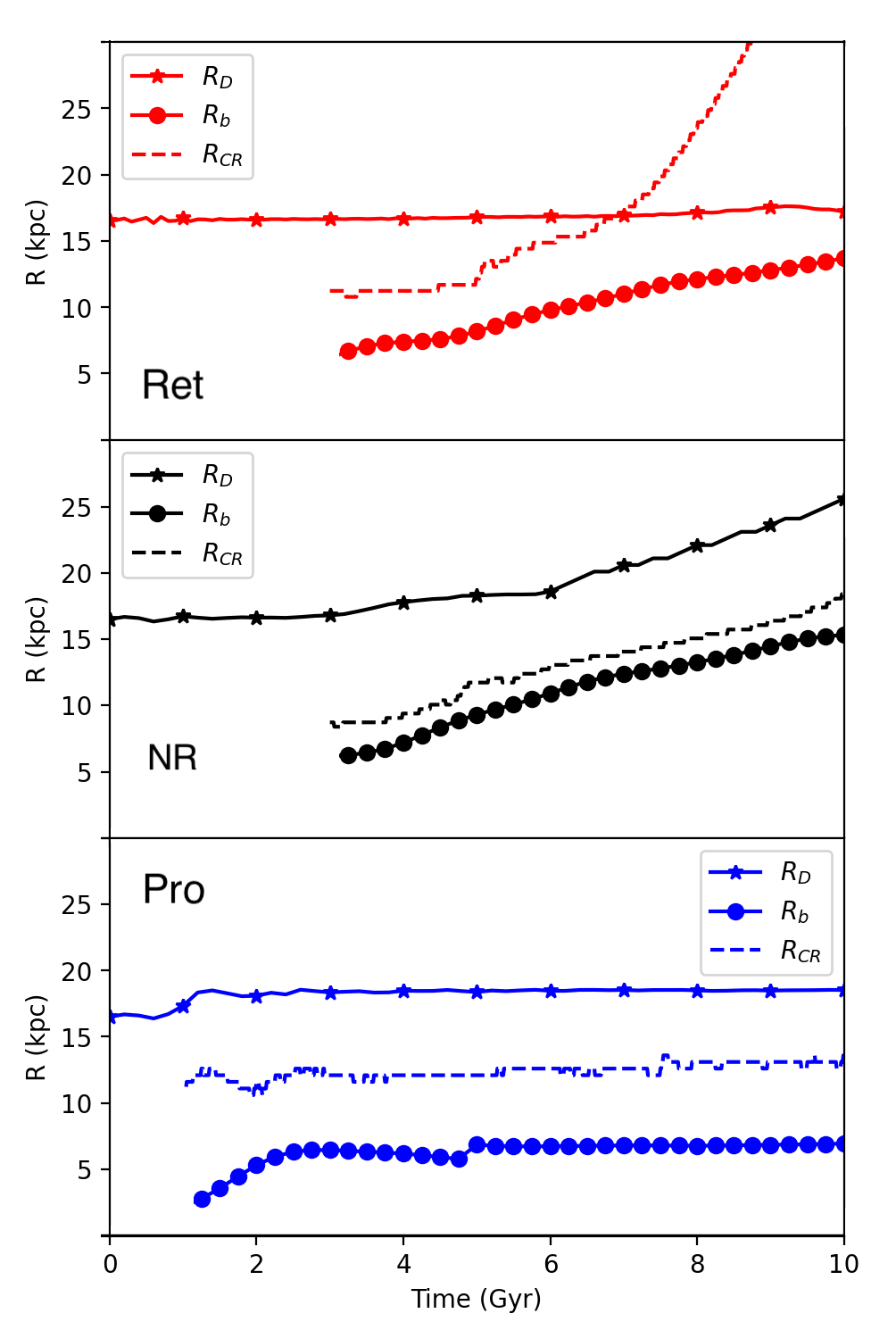}}
\caption{Evolution of stellar disk radius ($R_D$) in kpc (starred lines), the bar radius ($R_b$) in kpc (dotted lines) and the corotation ($R_{\rm{CR}}$) radius (dashed lines) for each model. The top panel shows the retrograde model (red), followed by the non-rotating model (black) and finally the prograde model (blue).}
\label{fig:CR_diskR}
\end{figure}

\begin{figure}
\centerline{
 \includegraphics[width=0.5\textwidth]{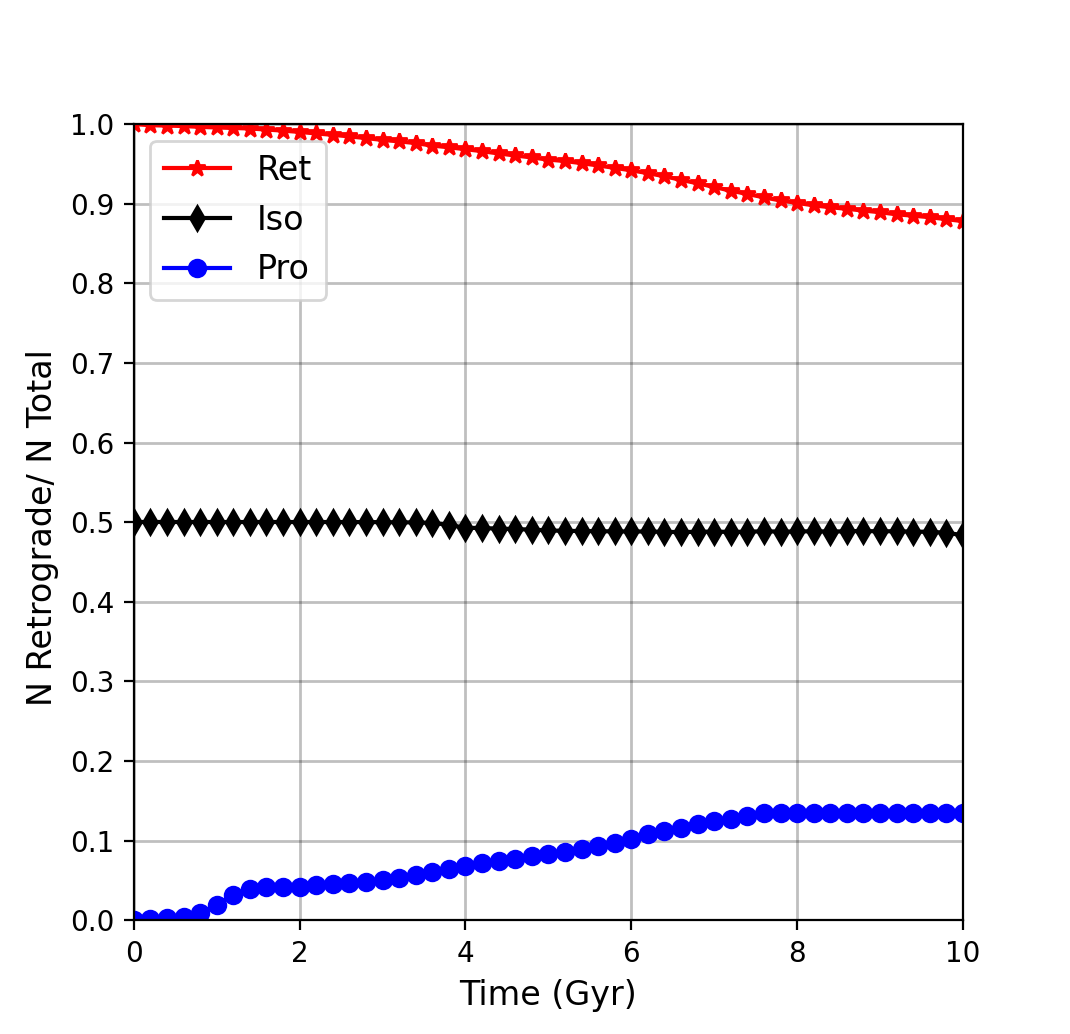}}
\caption{Fraction of dark matter halo orbits that are retrograde within $\lvert z \rvert \ltorder{10}$ kpc and $R \ltorder{30}$ kpc. Models are shown as (blue) circles for the prograde model, (black) diamonds for the non-rotating model, and (red) stars for the retrograde model.}
\label{fig:reversal}
\end{figure}

In Figure \ref{fig:CR_diskR} we plot the time evolution of the radius of the stellar disk, $R_D$, the length of the stellar bar, $R_b$, and the corotation radius, $R_{\rm{CR}}$, for each model. $R_D$ is the radius containing $97\%$ of the stellar material which is similar to the observable $R_{25}$. Bar length is given by the radius where the ellipticity of isodensity contours of stellar particles in the $xy-$plane drops to $85\%$ its maximum value \citep{marti06,coll19a}. $R_{\rm{CR}}$ is determined from the pattern speed of the bar, the radius at which $\Omega_b$ matches the circular velocity of the disk.

The bar in the prograde model {(bottom panel of Figure \ref{fig:CR_diskR})} remains relatively short throughout the simulation.
We have shown that prograde dark matter orbits are easily trapped by the stellar bar, forming a shadow dark matter bar (Figure~\ref{fig:wakes}).
This sweeping up of dark matter orbits into the bar potential reduces the torque on the stellar bar \citep{pet16}.
This is the reason that the bar pattern speed remains relatively high as seen in Figure \ref{fig:a2}.
The bar therefore does not slow down, trap more orbits, and grow in length. $R_{\rm{CR}}$ stops evolving in this model as the bar is not losing angular momentum and remains deep within the radius of the disk throughout the simulation.
In contrast, the non-rotating   model hosts a bar that continues to lengthen with time. $R_{\rm{CR}}$ moves out as the bar slows and the disk expands to larger radii.  $R_{\rm{CR}}$ still remains deep within the radius of the disk. The length of the bar in the non-rotating   model reaches much closer to $R_{\rm{CR}}$ than in the prograde model.

The retrograde model shows remarkably different evolution compared to the other models.
The disk remains at nearly the same radius for $10$ Gyr while the bar grows in length. The swift slowdown of the bar in this model moves $R_{\rm{CR}}$ far beyond the disk radius and prevents the disk from expanding due to resonant interactions with the bar.

\begin{figure*}
\centerline{
 \includegraphics[width=\textwidth]{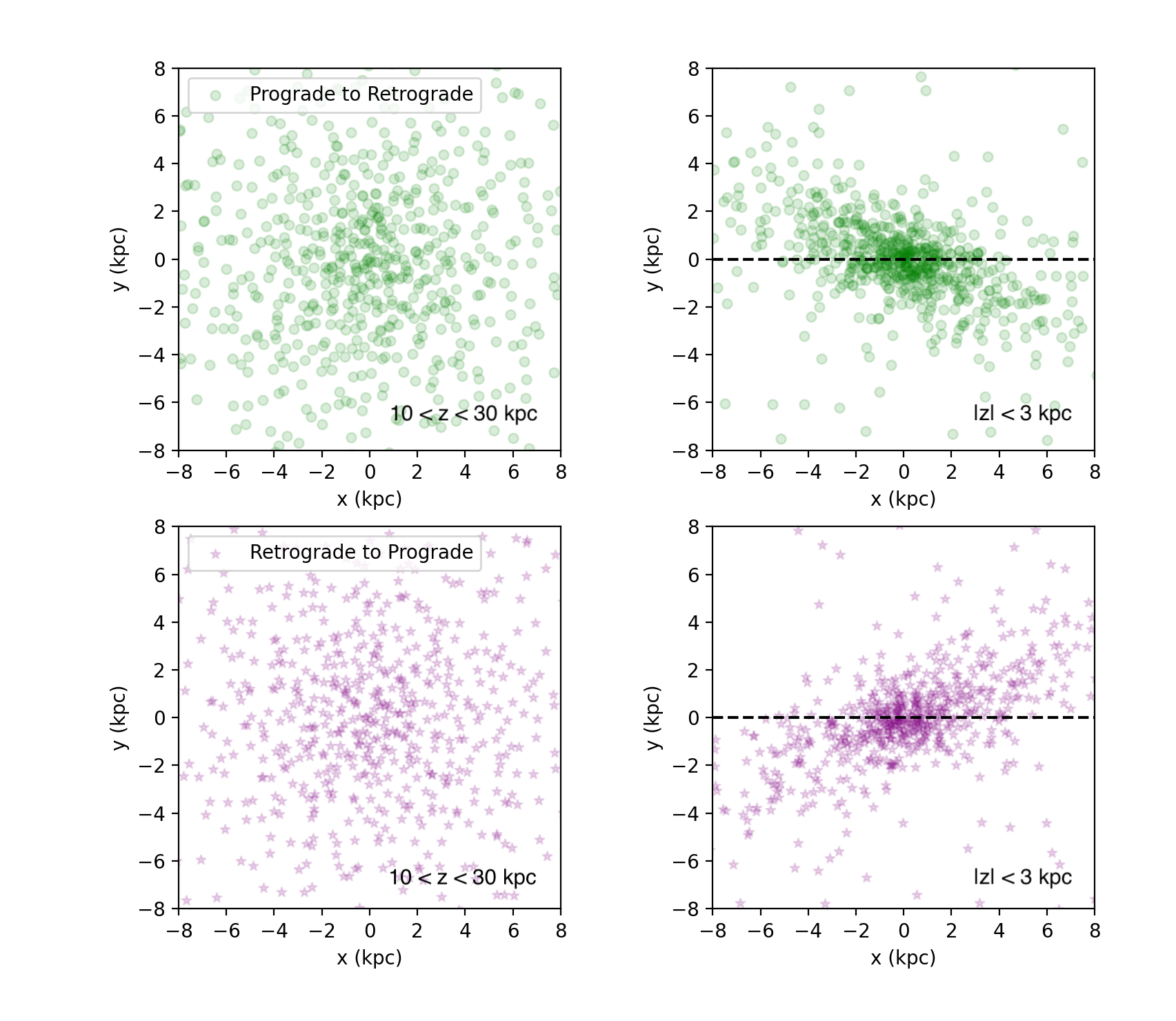}}
\caption{Instantaneous $xy$-positions of dark matter particles at the moment they reverse their orbital orientation, tracked over 10 Gyr in the non-rotating (NR) simulation. The left column shows reversals in the halo far from the disk ($10<z<30$ kpc) and the right column shows reversals in a slice of the dark matter halo containing the stellar bar ($\lvert z \rvert <3$ kpc). The top (bottom) two frames show particles that begin the simulation prograde (retrograde) and become retrograde (prograde). All frames were rotated to keep the stellar bar aligned with $y=0$ denoted here by the dashed line. The bar rotates in the clockwise direction. We find that halo orbits near the plane of the bar are preferentially reversed with respect to the bar's location. }
\label{fig:reversal_sample}
\end{figure*}

\subsection{Orbit Reversals}
\label{sec:reversal}

ln Figure \ref{fig:reversal} we plot the fraction of dark matter retrograde halo orbits within the region of influence of the bar for each model. This fraction decreases in the retrograde model and increases in the prograde model. The non-rotating model shows the fraction of retrograde orbits decreasing slightly over $10$ Gyr.

The prograde model shows an uptick in number of  retrograde orbits after $\sim 1$ Gyr. At this time, the bar is massive enough to torque prograde dark matter bar orbits which move ahead of the bar to retrograde orientations. In flipping dark matter orbits to retrograde orientations, the bar loses mass and gains angular momentum. This coincides with a drop in bar strength and a small increase in pattern speed {also around 1 Gyr} as seen in Figure~\ref{fig:a2}.
There is steady growth in the number of retrograde orbits until the bar buckles at $\sim7.5$ Gyr. Buckling reduces the strength of the bar which prevents additional reversals in this model.
The retrograde model shows a continuous reduction in the number of retrograde orbits (due to orbit reversals) as the bar grows in strength until the end of the simulation.
The non-rotating model has an equal number of retrograde and prograde halo orbits and both reversals should be at work at the same time. {However Figure \ref{fig:reversal} shows that the fraction of retrograde orbits decreases with time as more retrograde orbits flip to prograde orientation. We address the reason for this at the end of this section. }

{
To demonstrate the importance of the stellar bar in these orbital reversals, we track where they occur in the non-rotating model over 10 Gyr.
In Figure~\ref{fig:reversal_sample} we plot the instantaneous $xy$-position of the particles as they reverse orbital orientation, having first aligned the stellar bar along the $x$-axis.
In the left column, we do this in a slice of the halo far from the disk: $10 < z < 30$ kpc. We would expect that, far from the influence of the stellar bar, the orbit reversals would occur at random locations due simply to their chaotic dynamical nature. This is what we see.   
The top left panel shows the location in the $x/y-$plane where halo orbits have flipped from prograde to retrograde (green circles). Similarly, the bottom left panel shows the location of retrograde orbits that flip and maintain a prograde direction (purple stars). These panels show that, far from the disk, halo orbital reversals occur at random locations with respect to the stellar bar. }

{The right column of Figure~\ref{fig:reversal_sample} tells a different story. Here we show the instantaneous $xy$-position of the particles as they reverse orbital orientation in a slice of the dark matter halo containing the stellar bar ($\lvert z \rvert <3$ kpc). 

The reversals occur at distinct locations with respect to the bar.
The top right panel shows the locations where prograde halo orbits flip to retrograde orientation. The bar rotates in the clockwise direction.
As discussed in Section~\ref{halo-bar}, as low-inclination prograde orbits precess ahead of the bar, they can feel such a strong negative torque that they can reverse their angular momentum direction. Hence, we find that prograde-retrograde orbit reversals occur primarily where orbits lead the stellar bar.  In this simulation, the average position of a reversing prograde-retrograde particle is at an azimuthal angle $\langle \phi \rangle \approx 37^\circ$ ahead of the bar.}

{The bottom right panel shows the opposite type of reversal. While the stellar bar rotates in the clockwise direction, low-inclination retrograde orbits precess in the counter-clockwise direction. After they speed through the stellar bar, these orbits feel a negative torque that slows them down. If the torque is sufficiently strong, the orbit can reverse its orientation. Hence retrograde-prograde orbit reversals occur primarily where orbits lag behind the stellar bar. In this simulation, the average position of a reversing particle is at an azimuthal angle $\langle \phi \rangle \approx 28^\circ$ behind the bar.}

\begin{figure}
\centerline{
 \includegraphics[width=0.5\textwidth]{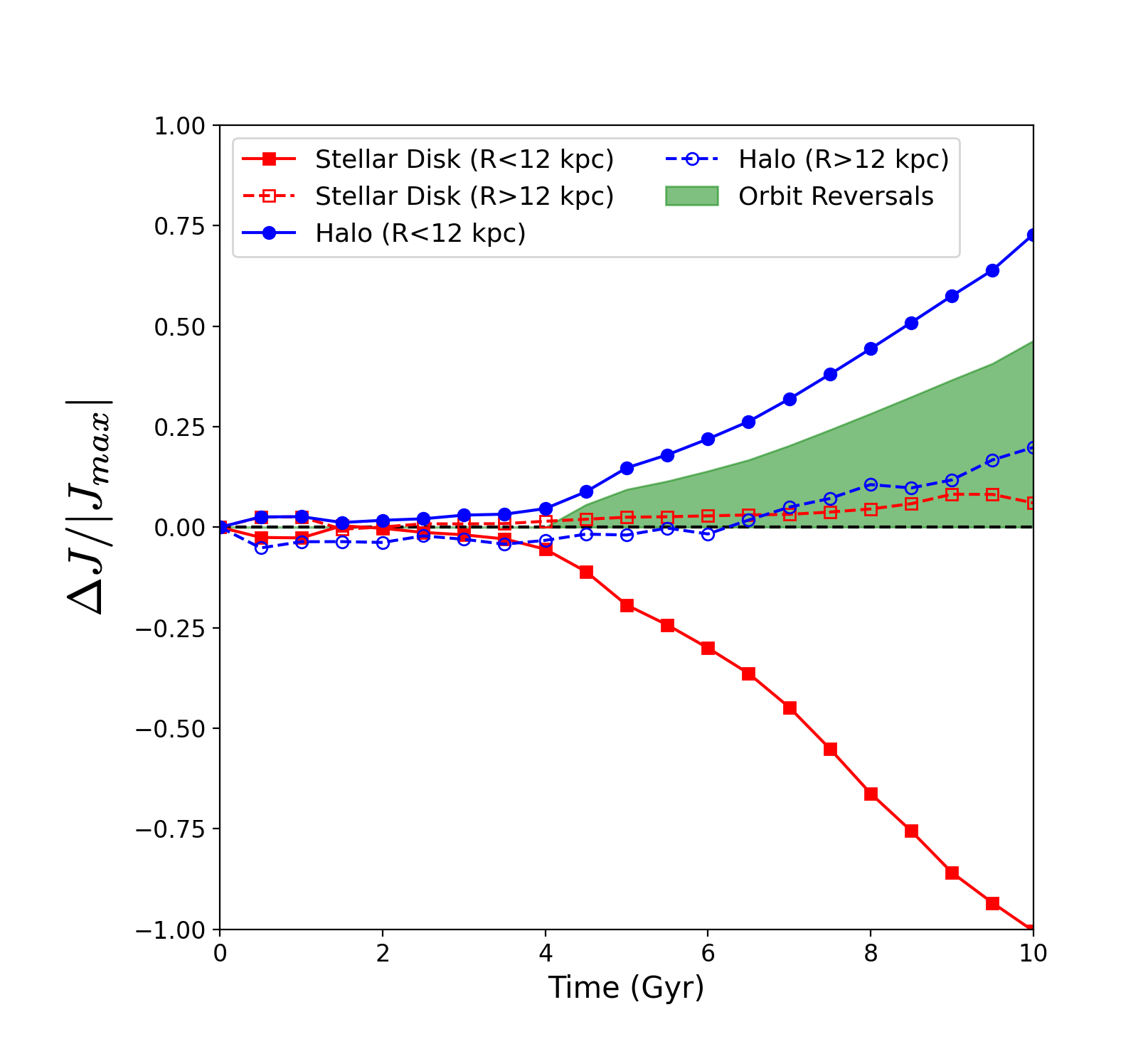}}
\caption{
Time evolution of change in angular momentum, $\Delta J=J(t) - J(t=0)$ normalized to the maximum amount ($J_{max}$) moved by the system. The measurement is made for four discrete regions, inside and outside $R=12$ kpc for each disk and dark matter halo. This plot is taken at a different $z-$limit than Figure \ref{fig:jz2} and includes an additional shaded (green) region showing the angular momentum gained by retrograde dark matter particles that undergo bar-driven orbit reversals.
The disk inside (outside) $R=12$ kpc are the filled (empty) squares. The dark matter halo inside (outside) $R=12$ kpc are the filled (empty) circles.}
\label{fig:reversalJ}
\end{figure}

{We have shown that the stellar bar plays an important role in reversing the orbits of low inclination dark matter halo particles. In Figure~\ref{fig:reversalJ} we quantify the contribution of these orbit reversals to the decrease in angular momentum and swift slowdown of the stellar bar in the retrograde model.
We select dark matter halo particles within $R<12$ kpc and $\lvert z \rvert < 6$  kpc that have a prograde orientation at $t=10$ Gyr and follow their angular momentum backwards with time.
While reversals happen throughout the halo (i.e., left panel of Figure \ref{fig:reversal_sample}), orbits reversals due to the stellar bar appear only in the region of influence of the stellar bar. The total change in angular momentum of these orbits is shown by the shaded green region in Figure \ref{fig:reversalJ}. This plot is similar to the top panel of Figure \ref{fig:jz2} but with a smaller vertical limit to limit the measurement to resonant, bar-driven angular momentum transfer. Bar-driven orbital reversals account for more than half of the angular momentum transfer between the inner disk and the inner halo in the retrograde model.
}

{The torque experienced by a stationary orbit in the presence of a uniformly-rotating m=2 mode should cancel to zero in one precession period\footnote{Here we define a precession period as the timescale over which the major axis of an orbit precesses by $2\pi$ radians in the orbital plane.}. Hence we would expect the retrograde dark matter wake to lie orthogonal to the bar. This is not the case in the top panel of Figure~\ref{fig:wakes} however. This figure shows that the surface density contours of the dark matter wake, averaged across $\sim0.1$ Gyr, are slightly offset from perpendicular. This is due to the bar-driven orbit reversals. As we see in Figure~\ref{fig:reversal_sample}, reversals preferentially occur {\it behind} the bar. These orbits leave the retrograde population, creating an under-density {\it ahead} of the bar. When averaged over time, this skews the surface density contours of the wake towards the position of peak retrograde orbit reversals.}

The preferred location of reversals is similarly explained for prograde orbits. As seen in Figure \ref{fig:reversal_sample}, orbits flip from prograde to retrograde ahead of the bar. These reversals remove negative angular momentum from the bar. As a result, the stellar bar in the prograde halo gains angular momentum.  However it loses angular momentum to the outer disk and halo. The resulting bar pattern speed is thus nearly constant until buckling (Figure \ref{fig:a2}) as the angular momentum losses and gains balance (bottom panel of Figure \ref{fig:jz2}). 

{A non-rotating dark matter halo has an equal number of prograde and retrograde orbits. However the effects on the stellar bar from each of these populations do not cancel each other out.
The bar has a higher rate of gravitational encounters with low inclination retrograde orbits as they differentially precess with respect to one another. At $t = 10$ Gyr, the percentage of prograde dark matter orbits in our non-rotating model is 53\%: there have been more retrograde-prograde orbit reversals. The bar slows down (Figure \ref{fig:a2}) albeit not as fast as in the retrograde model. As long as the bar keeps growing and encountering more retrograde orbits, it will continue to decrease in pattern speed. A similar mechanism in the near-Keplerian potential of a galactic nucleus is described in \citet{Madigan2012}. }

\section{Observational Consequences}
\label{obs-cons}

\subsection{Baryonic observables}

The first galaxy classification system was developed by Edwin Hubble who mapped galaxies onto a fork diagram \citep{hubble26,hubble36}. The Hubble sequence divides galaxies into ellipticals and spirals. The spiral sequence is further divided into two branches, one for barred spirals and one for unbarred spirals. In Figure \ref{fig:dens} we show that we can create both lanes of the spiral portion of the Hubble sequence from an isolated simulation of an identical disk simply by varying the fraction of prograde/retrograde orbits in the dark matter halo. Prograde halos form large dark matter bars which lengthens the timescale of the buckling instability. This buckling weakens the bar significantly (Figure \ref{fig:a2}) and the resulting disk is too hot to undergo a second bar instability. This effect is described in detail in \citet{coll18}.

Halo spin affects the early stages of disk evolution, with a higher fraction of prograde orbits shortening the time scale of bar formation. As the presence of bars compresses gas and accelerates star formation in the early stages of bar evolution \citep{sellwood14,vera16, kim17} we would expect galaxies of prograde halo spin to show different star formation rates, color, and chemical composition when compared to galaxies with non-rotating or retrograde spin. The James Webb Space Telescope, which at the time of this writing is expected to launch in 2021, will see in high resolution the era of galaxy formation for the first time. Our understanding of how dark matter kinematics effects early disk evolution may be useful in interpreting the newly acquired images of galaxy assembly \citep{wind06,kali18}.

\begin{figure}
\centerline{
 \includegraphics[width=0.5\textwidth]{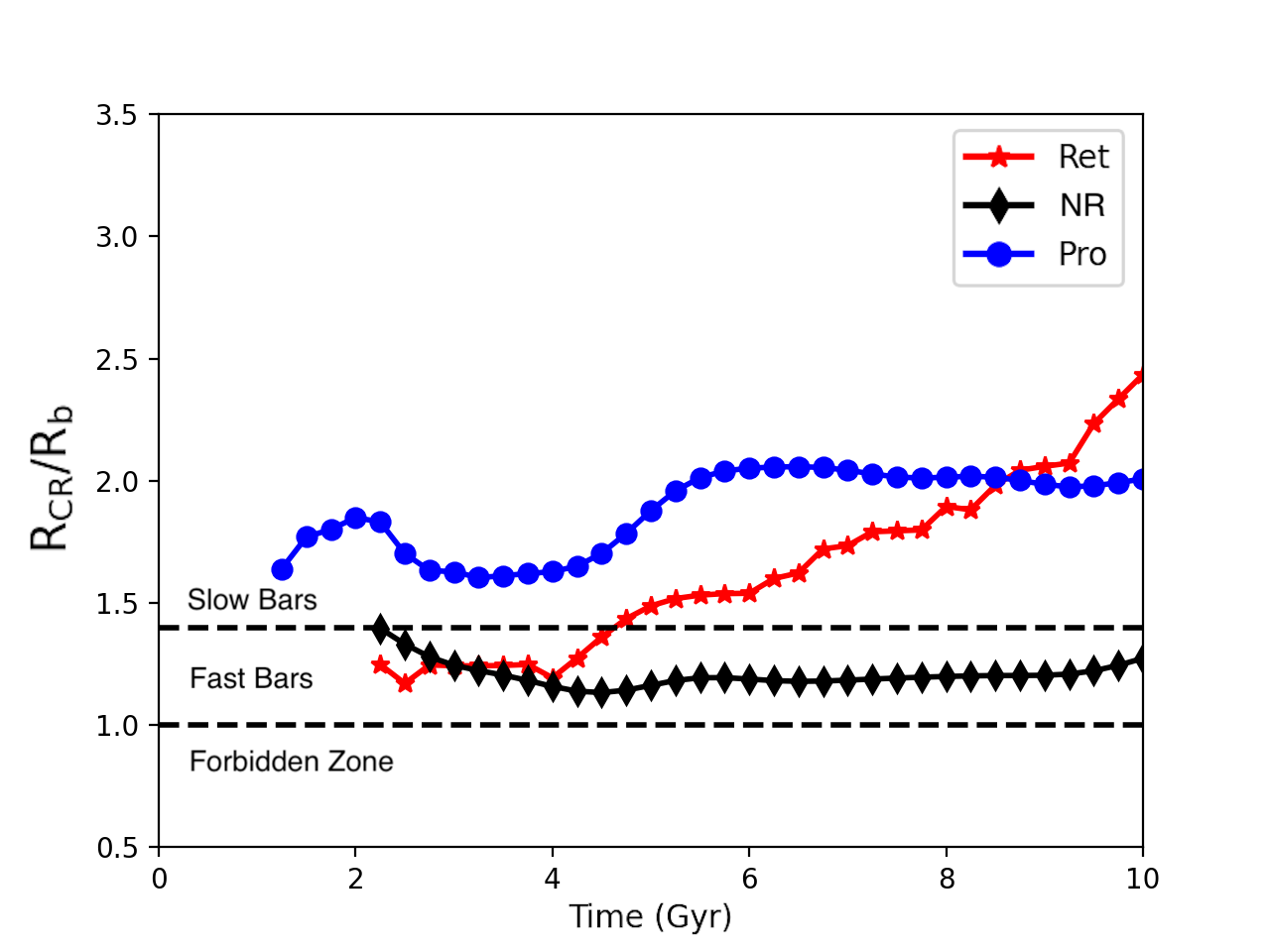}}
\caption{Time evolution of the ratio of the corotation radius ($R_{\rm{CR}}$) to bar radius ($R_b$). The dashed lines show the limits for `fast bars' where $R_{CR}/R_b = 1.2 \pm 0.2$. Models are shown as (blue) circles for the prograde model, (black) diamonds for the non-rotating model, and (red) stars for the retrograde model.}
\label{fig:cr-rb}
\end{figure}

Halo-bar coupling has a strong influence on the way angular momentum is transferred throughout the galaxy, affecting observables such as bar and disk morphology (Figures \ref{fig:dens} and \ref{fig:CR_diskR}.)
An observable property of barred disk galaxies is the ratio of the corotation resonance, $R_{\rm{CR}}$, to the length of the stellar bar, $R_b$. Bars that fall into the narrow range of $R_{\rm{CR}}/R_b = 1.2 \pm 0.2$ are called `fast' bars (as the bar pattern speed is near the angular speed of the stars at that radius), and bars that fall above this range are `slow' bars \citep{teuben85,athan92,marti06}.
We plot this ratio for each model in Figure \ref{fig:cr-rb}.
The ratio remains fairly constant in the non-rotating model, consistent with recent modeling of a slowing bar by \citet{Chiba2021}. This is in contrast with the other two models, particularly the retrograde model. In this case, the co-rotation radius evolves much faster in time than the bar radius.
The non-rotating model lies in the regime of `fast' bars while the retrograde and prograde models \deleted{evolve} \added{are} well beyond this range and end up as `slow' bars. \added{Figure \ref{fig:cr-rb} shows $R_{\rm{CR}}/R_b$ evolving for the Ret model, while the bars in the Pro and NR models change only slightly from their initial value. This result indicates that strongly torqued bars can slow down so quickly they will move from the `fast' to `slow' regime.} The nomenclature of `fast' versus `slow' bars is a misnomer in the case of the prograde model as the bar has the fastest pattern speed out of all the models (Figure \ref{fig:a2}). 
The vast majority of bars have been interpreted observationally to be fast \citep[e.g.,][]{aguerri03,Chemin09,corsini11}.
However, recent results from the SDSS-IV MaNGA survey may change this picture. After careful characterization of error sources, \citet{manga2020} find 7 slow bars in a sample size of 18 barred galaxies. We have shown that strong rotation in either direction will push the stellar bar from the fast bar regime to the slow one, providing a simple dynamical explanation for the existence of slow bars.

\begin{figure}
\centerline{
 \includegraphics[width=0.5\textwidth]{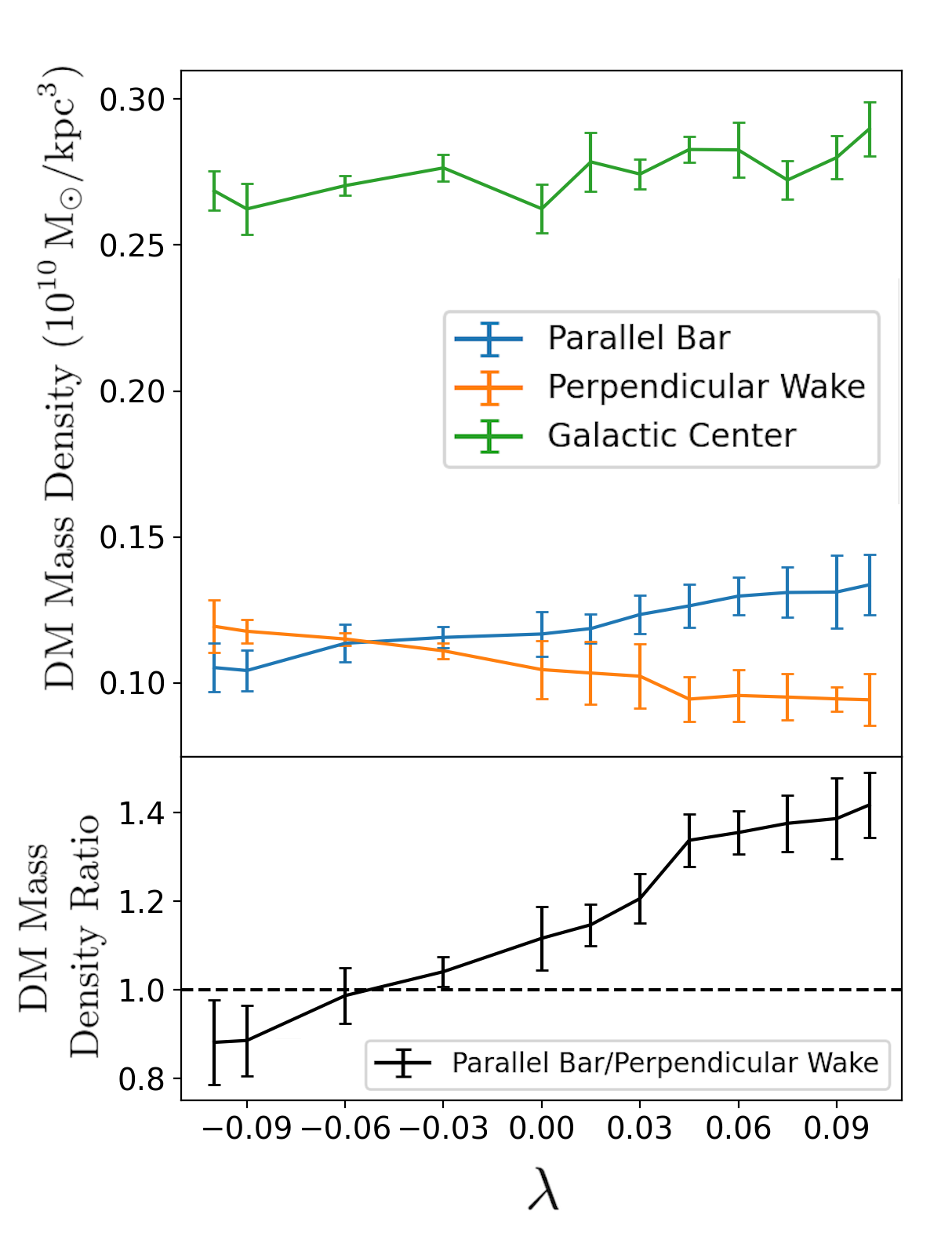}}
\caption{Top: Mass density of dark matter (DM) measured in a sphere of radius $0.5$ kpc at three points in each galaxy of varying dark matter halo spin, $\lambda$.
Bottom: Ratio of the parallel dark matter bar mass density over the perpendicular dark matter wake mass density. Measurements are taken when the stellar bar is at maximum strength late in the evolution in each model.
We plot average values using data from five nearby snapshots and error bars show the standard deviation.}
\label{fig:dark matter_column_den}
\end{figure}

\subsection{Indirect Dark Matter Detection}

Many candidate dark matter particles should annihilate and/or decay resulting in observable electromagnetic radiation. For a review of indirect detection of particle dark matter see \citet{Gaskins_16}.
In this paper we show how the coupling of the dark matter halo to the stellar bar results in local regions of higher dark matter density within the centers of galaxies.
{In the case of a strong perpendicular dark matter wake, this region of high  dark matter density appears in a region of relative low baryon density. This is very clear in the top panel of Figure \ref{fig:dm-angle}.}

In Figure \ref{fig:dark matter_column_den} we plot the mass density of dark matter at three locations in each model.
The three measurements are made in a sphere of radius $0.5$ kpc at 1) the galactic center, 2) parallel to the stellar bar at a radius of $R_b/2$, and 3) perpendicular to the stellar bar at a radius of $R_b/2$.
Each quantity is computed when the stellar bar is at maximum strength late in the evolution for each model. We compute the average using data from five nearby snapshots; error bars show the standard deviation.
We have included the results from models with spins in between the fully prograde and retrograde models for illustration.

The top panel of Figure \ref{fig:dark matter_column_den} shows how dark matter mass density varies with spin of the halo, $\lambda$.
Prograde orbits become trapped in the stellar bar which leads to an increase in the parallel dark matter bar density as $\lambda$ increases in the positive, prograde direction.
The reverse occurs with retrograde orbits as they increase the mass density of dark matter in the perpendicular wake.
This figure shows that the over-density of dark matter parallel and perpendicular to the stellar bar can be significant, albeit lower, compared to the dark matter density at the galactic center. In maximally prograde spinning halos, the parallel dark matter bar can reach $\sim40\%$ the density at the galactic center.
The bottom panel of Figure \ref{fig:dark matter_column_den} shows the ratio of the mass density in the parallel dark matter bar versus perpendicular wake. This ratio increases linearly with $\lambda$.

\begin{figure}
\centerline{
 \includegraphics[width=0.5\textwidth]{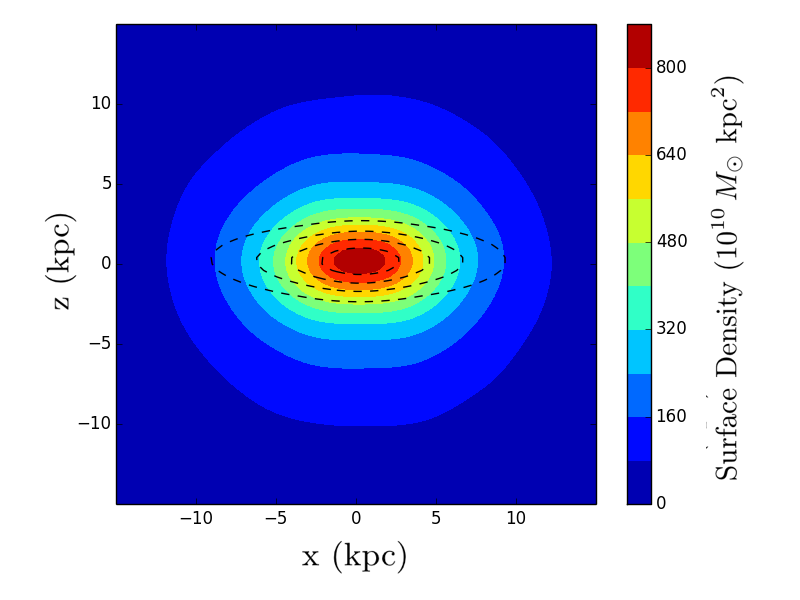}}
\caption{Surface density contour plot of an edge-on slice ($|y|<2$ kpc) of the dark matter halo in prograde model at $t=7.0$ Gyr. The dark matter particle values are marked by rainbow contours. Stellar particles are indicated by the black dashed contours. The increasing contours of the stellar bar match the levels on the first four dark matter halo contours but are larger by an order of magnitude.}
\label{fig:dark matter_height}
\end{figure}

In prograde dark matter halos, dark matter orbits become trapped in the stellar bar. Their inclinations are higher than those in the stellar disk which leads to a dark matter bar that is thicker in vertical extent than the stellar bar.
We show an edge-on view of the galaxy in the prograde model at $t=7.0$ Gyr in Figure \ref{fig:dark matter_height}.
Rainbow contours show the surface density contours of dark matter particles. Dashed lines show the equivalent for stellar particles.
{The scale height, measured as the vertical distance over which the density falls by a factor $e$, of the stellar bar is $0.8$ kpc while the scale height of the dark matter bar is $2.4$ kpc.}

\subsection{The Milky Way Galaxy}

The Milky Way has gone $\sim10$ Gyr without a major merger; see \citet{Bland16} for a comprehensive review.
Cosmological simulations predict that the inner dark matter halo of the Milky Way is consolidated at this time, while the accretion of satellites continues to impact the outer halo 
\citep{Bullock05,Pillepich14}.
Therefore, the inner Milky Way halo has had ample opportunity for halo-bar coupling.  A recent study of the metallicity gradient found in Solar neighborhood stars by \citet{chiba2021-2} provides direct evidence of bar slow down which in turn supports the existence of a Standard Model Halo which interacts with the Milky Way bar via resonant angular momentum exchange.  
{As we have shown, this coupling should} result in the production of a parallel dark matter bar and a perpendicular dark matter wake, the strength of which will depend on the {density distribution and} the spin ($\lambda$) of the inner Milky Way halo.

\citet{Wegg15} measure the bar angle, defined as the angle of the bar's major axis to the Sun line-of-sight ($\Theta_b$) to be between $28^{\circ}-33^{\circ}$. Therefore we predict the perpendicular dark matter wake angle to be $55^{\circ} \lesssim \Theta_{\perp \rm{DM}} \lesssim 60^{\circ}$. The end of the nearside of the perpendicular dark matter wake should have a galactic longitude of $l \approx 323^\circ$. This could provide an opportunity to measure a dark matter over-density far from the bright Galactic center. 

\section{Conclusions}
\label{sec:conclusion}

In this paper we use high-resolution numerical simulations of galaxies embedded in rotating dark matter halos to demonstrate the coupling of dark matter orbits with stellar bars. Our results are as follows:

\begin{itemize}

    \item The inner dark matter halo couples to the stellar bar. Low inclination prograde dark matter orbits join the bar, increasing its mass and reducing the torque it would otherwise experience (first noted by \citet{pet16}). Low inclination retrograde dark matter orbits slow as they precess behind the bar, forming a perpendicular dark matter wake. Parallel and perpendicular dark matter over-densities develop in all halos hosting a stellar bar, even in those with zero net angular momentum and spin parameter.
    These dark matter over-densities provide a novel space to look for dark matter annihilation or decay signals.

\item The Milky Way Galaxy should host both a parallel dark matter bar and a perpendicular dark matter wake, the near-side of which should extend from Galactic center to a galactic longitude of $l \approx 323^\circ$.

\item The time-averaged retrograde dark matter wake is slightly misaligned from perpendicular. This is due to bar-driven orbit reversals which for retrograde orbits occur behind the bar and skew the surface density contours of the wake in their direction.

 \item Counter-rotating, {\it stellar} bars were first explored by \citet{frie96} and a perpendicular dark matter wake was first observed in the {\it dark matter} distribution in simulations by \citet{coll19b}. To our knowledge, this is the first time that their formation mechanism has been explained.

\item Orbital reversals are an important channel of angular momentum transfer between the stellar bar and the dark matter halo. Even in initially non-rotating dark matter halos, bars slow down due to a bar-driven reversals of retrograde dark matter orbits. 

    \item  Retrograde orbit reversals exert a negative torque on the stellar bar, leading to its elongation and decrease in pattern speed. In strongly retrograde halos, this reduces angular momentum transfer to the outer disk by moving the corotation radius well beyond the outer disk radius.
     \added{The rapidly increasing corotation radius outpaces the growth of the bar in our retrograde model. As a result, the bar in this system evolves from the `fast' to the `slow' bar regime over Gyr timescales. }

    \item The varying morphologies that appear in our isolated, initially identical disks show that baryons make visible the underlying kinematics of the dark matter halo.  The work of measuring the dark matter kinematics of the Milky Way halo via stellar observations has already begun with the GAIA collaboration \citep{gaia}. We have shown how the dark matter bar mass is dependent on halo spin (Figure \ref{fig:dark matter_column_den}). Future measurements of dark matter annihilation signals either in the parallel dark matter bar or the dark matter wake in our own Galaxy could provide an alternate way of indirectly measuring spin in the Milky Way's dark matter halo.

\end{itemize}

In future, we will explore the secular change in dark matter halo shape (from spherical to an oblate spheroid) in the presence of a strong stellar bar and its effect on dark matter annihilation/decay signals in the Galaxy. The work presented here predicts that the changing potential of dark matter will produce an asymmetric signal near the Galactic center due to the strong bar in the Milky Way.  Future work will investigate whether the change in potential is enough to produce an asymmetry in annihilation/decay signals that is observable from Earth.
Additionally, as the halo-bar coupling has far-reaching dynamical consequences for the outer stellar disk, we will investigate how the spin of the dark matter halo affects spiral arm regeneration and the dark matter density distribution within them.

\section*{Acknowledgements}
Many thanks to Gongjie Li and Katelin Schutz for helpful conversations.
A.M. gratefully acknowledges support from the David and Lucile Packard Foundation.
This work was supported by a NASA Astrophysics Theory Program under grant NNX17AK44G.  We use the RMACC Summit supercomputer, which is supported by the National Science Foundation (awards ACI-1532235 and ACI-1532236), the University of Colorado Boulder, and Colorado State University. The Summit supercomputer is a joint effort of the University of Colorado Boulder and Colorado State University.

\software{GIZMO (\cite{hop15})}

\section*{Data Availability}
The data underlying this article will be shared on reasonable request to the corresponding author.

\bibliographystyle{mn}
\bibliography{main}



\end{document}